\documentclass[aps,pra,twocolumn,groupedaddress,showpacs,floatfix]{revtex4-1}

\usepackage{times,amssymb,amsmath}
\usepackage{array}
\usepackage{ifpdf}
\usepackage{txfonts} 
\usepackage{graphicx}

\mathchardef\mhyphen="2D

\renewcommand{\Im}{\text{Im}\,}
\usepackage{cleveref}

\newcommand{\bsube}{\begin{subequations}}
\newcommand{\esube}{\end{subequations}}
\newcommand{\be}{\begin{equation}}
\newcommand{\ee}{\end{equation}}
\newcommand{\bea}{\begin{eqnarray}}
\newcommand{\eea}{\end{eqnarray}}

\newcommand{\nn} {\nonumber}

\begin{document}


\title{Taming singularities of the diagrammatic many-body perturbation theory}
\author{Y. Pavlyukh}
\email[]{yaroslav.pavlyukh@physik.uni-halle.de}
\author{J. Berakdar}
\affiliation{Institut f\"{u}r Physik, Martin-Luther-Universit\"{a}t
  Halle-Wittenberg, 06120 Halle, Germany}
\author{A. Rubio}
\affiliation{Max Planck Institute for the Structure and Dynamics of Matter and 
Center for Free-Electron Laser Science and Department of Physics, Luruper Chaussee 149, 
22761 Hamburg, Germany}
\affiliation{Nano-Bio Spectroscopy Group and ETSF Scientific Development Centre,
 Dpto. de F{\'i}sica de Materiales, Universidad del Pa{\'i}s Vasco,
 CFM CSIC-UPV/EHU-MPC and DIPC, E-20018 San Sebasti{\'a}n, Spain}
\begin{abstract}
In a typical scenario the diagrammatic many-body perturbation theory generates asymptotic
series.  Despite non-convergence, the asymptotic expansions are useful when truncated to a
finite number of terms. This is the reason for popularity of leading-order methods such as
$GW$ approximation in condensed matter, molecular and atomic physics.  Emerging
higher-order implementations suffer from the appearance of nonsimple poles in the
frequency-dependent Green's functions and negative spectral densities making
self-consistent determination of the electronic structure impossible.  Here a method based
on the Pad{\'e} approximation for overcomming these difficulties is proposed and applied
to the Hamiltonian describing a core electron coupled to a single plasmonic excitation. By
solving the model purely diagrammatically, expressing the self-energy in terms of
combinatorics of chord diagrams, and regularizing the diverging perturbative expansions
using the Pad{\'e} approximation the spectral function is determined self-consistently
using 3111 diagrams up to the sixth order.
\end{abstract}
\maketitle
Introduction of the Green's function methods to electronic structure calculations is the
most prominent achievement of the field-theoretic
methods~\cite{nozieres_theory_1999,gogolin_bosonization_1998,giuliani_quantum_2005} on par
with the density functional theory having immediate technological
applications~\cite{dreizler_density_1990,onida_electronic_2002}. Even in the lowest
(beyond the mean field) order one obtains significant improvements of e.\,g. the band gap
through the correlational shifts ($\Delta$). Including higher-order diagrams (vertex
corrections) is numerically demanding. However, there are more fundamental obstacles on the
way arising from dealing with \emph{diverging series} as the following consideration
illustrates.
\paragraph*{Pad{\'e} approximation\hspace{-1em}}---
\begin{figure}[b!]
\includegraphics[width=\columnwidth]{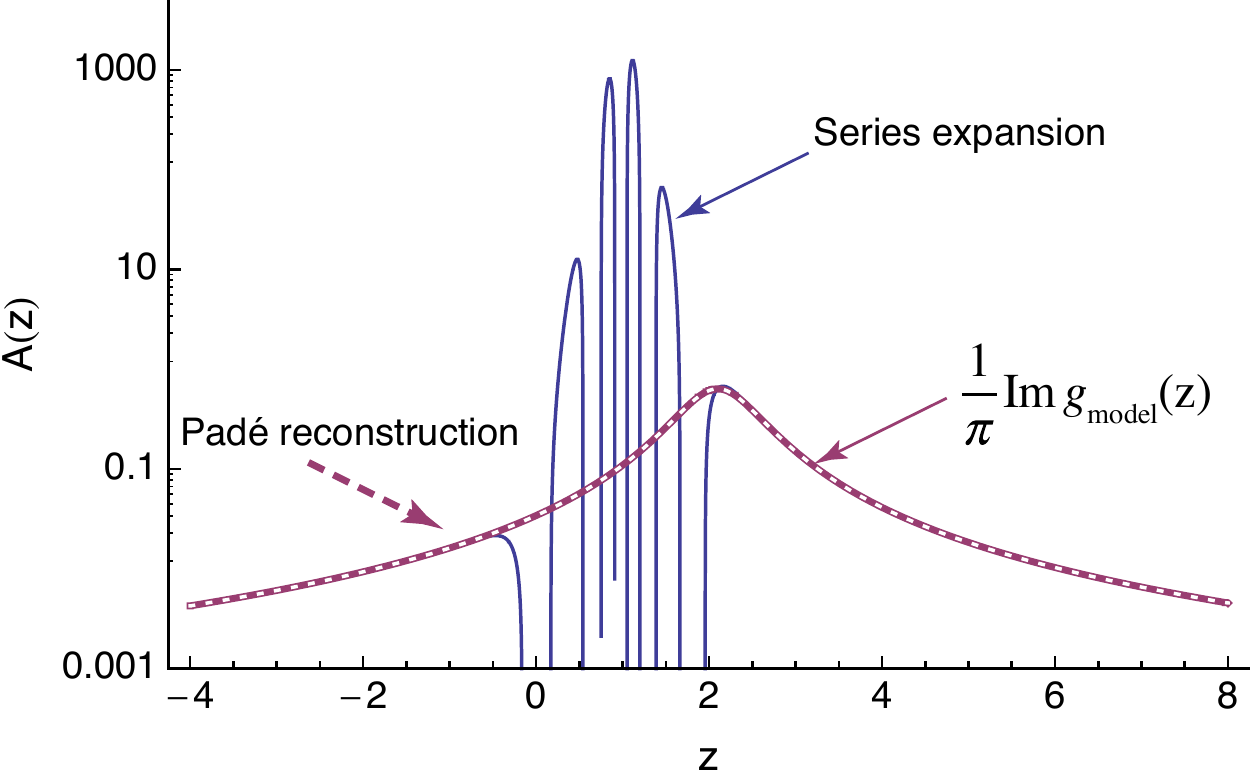}
\caption{Reconstruction of $g_{\textrm{model}}(z)$ from its series
  expansion~(\ref{eq:mod}) in terms of $g_{\textrm{model}}^{(0)}(z)$ using the Pad{\'e}
  approximation. Parameters are as follows: $\Delta=1.1$, $\epsilon=1$, $\eta=0.5$. The
  series expansion~(\ref{eq:mod}) is restricted at $n_\textrm{max}=10$, and the Pad{\'e}
  approximation is applied at the point $z=6$. Notice that the original (magenta) and
  reconstructed (white) densities of states are practically indistinguishable.
\label{fig:Pade}}
\end{figure}
Let $g_{\textrm{model}}(z)=1/(z-\epsilon -\Delta- i\eta)$ be a model Green's function
(GF), and $\Delta$ the energy shift due to some interaction. It can be expanded in terms
of the non-interacting GF $g_{\textrm{model}}^{(0)}(z)=1/(z-\epsilon -i\eta)$ as geometric
series:
\be
g_{\textrm{model}}(z)=\sum_{n=0}^\infty\frac{\Delta^n}{(z-\epsilon -i\eta)^{n+1}}.
\label{eq:mod}
\ee 
The series expansion behaves oscillatory in the vicinity of the pole and approaches the
original function at large $z$, i.\,e., for $\left|\frac{\Delta}{z-\epsilon
  -i\eta}\right|<1$. Nonetheless, a sensible spectral function, $A(z)=\frac1{\pi}\Im
g(z)$, in the domain of interest can be reconstructed by using the Pad{\'e}
approximation. The procedure is outlined at Fig.~\ref{fig:Pade} where the original
function $g_\text{model}(z)$, series expansion~(\ref{eq:mod}) and the Pad{\'e}
reconstruction are shown.  The Pad{\'e} approximation allows to obtain very accurate
values also in the domain where the series~(\ref{eq:mod}) is diverging. The method works
so well here because it is known in advance that GF consists of one pole only and this
fact is used for the reconstruction (according to the exact form of
$g_{\textrm{model}}(z)$ we use the $[0/1]$ approximant~\footnote{The Pad{\'e} approximation
  has form of rational function (denoted as $[M/N]$) with $M+N+1$ coefficients, $M$, $N$
  are the orders of the numerator and denominator, respectively.}).  For realistic
calculations we do not have this knowledge and have to rely on some additional assumptions
about the analytic structure of the Green's function. As an illustration let us consider
the electron-boson Hamiltonian --- an ubiquitous in condensed matter physics model.

\paragraph*{Model specification and known results\hspace{-1em}}---
Consider a set of fermionic and bosonic quantum numbers and the associated creation and
annihilation operators with standard commutation rules:
\be
\big[c_a,\,c_b^\dagger\big]_{+}=\delta_{ab},\quad
\big[a_i,\,a_j^\dagger\big]_{-}=\delta_{ij}.
\ee
The model becomes non-trivial when a coupling between fermionic and bosonic degrees of
freedom is introduced 
$\mathcal H_I=\sum_{ab}\sum_i \Gamma_{ab}^ic_a^\dagger c_b a_i+\text{H.c.}$
This very general model covers various physical scenarios: (i) interaction of electrons in
solids with \emph{real} bosonic excitation such as phonons forming the basis of the
polaron model (Sec 4.3 of Mahan~\cite{mahan_many-particle_2000}), novel applications
include quantum dots coupled to nanomechanical oscillators~\cite{tahir_novel_2014};
(ii)~electronic excitations such as plasmons under some assumptions mediate the
electron-electron interaction. This scenario was first introduced in the work of
Lundqvist~\cite{lundqvist_characteristic_1969} who considered coupling of the deep hole to
plasmonic excitations in metals with well known analytic
solution~\cite{langreth_singularities_1970,almbladh_comments_1978,cini_exactly_1988}. Another
prominent example is the photoemission process where the photoelectron interacts with the
density fluctuations of the target system~\cite{hedin_transition_1998}; (iii) auxiliary
bosonic degrees of freedom is a mathematical trick used to treat a pure electronic
Hamiltonian such as the mixed-valence Hamiltonian, i.\,e. large-$U$ Anderson model
(\emph{slave-boson} approach)~\cite{coleman_new_1984}.

Besides the form of the Hamiltonian, it is the notion of the ground state that determines
the diagrammatic structure of the model. For instance, the no-hole state is of relevance
for the x-ray absorption in the Lundqvist model, while for the photoemission one considers
a state with exactly one deep hole.  At variance, the ground state of the large-$U$
Anderson model is determined as a state in which the sum of boson and fermion occupation
numbers at each site is unity.  For this two-component fermionic model very different
diagrams (non-crossing approximation) are relevant~\cite{wingreen_anderson_1994}.

Consider the electron-boson Hamiltonian in its simplest form:
\begin{equation}
\mathcal H=\epsilon c^\dagger c+cc^\dagger \gamma(a+a^\dagger)+\Omega a^\dagger a,\label{eq:H}
\end{equation}
where $c$ is the creation operator of the deep hole with energy $\epsilon$, $a^\dagger$ is
the bosonic creation operator of the plasmon with the energy $\Omega$.  The facts that
there might be several kind of fermions as in the mixed-valence impurity model or the
plasmon dispersion are neglected here. However, the generalization to the latter case is
possible and will be commented on after the presentation of the diagrammatic solution. The
Hamiltonian~(\ref{eq:H}) is quite versatile and is applicable to other scenarios such as
resonant-tunneling through a single level coupled to wide-band
phonons~\cite{wingreen_resonant_1988}. Remarkably, also the two particle GF can be found
analytically~\cite{wingreen_inelastic_1989}, the model can be solved at finite
temperatures, and its non-equilibrium properties have also been studied
thoroughly~\cite{dash_nonequilibrium_2010,ness_gw_2011}.

\begin{figure}[t!]
\includegraphics[width=0.9\columnwidth]{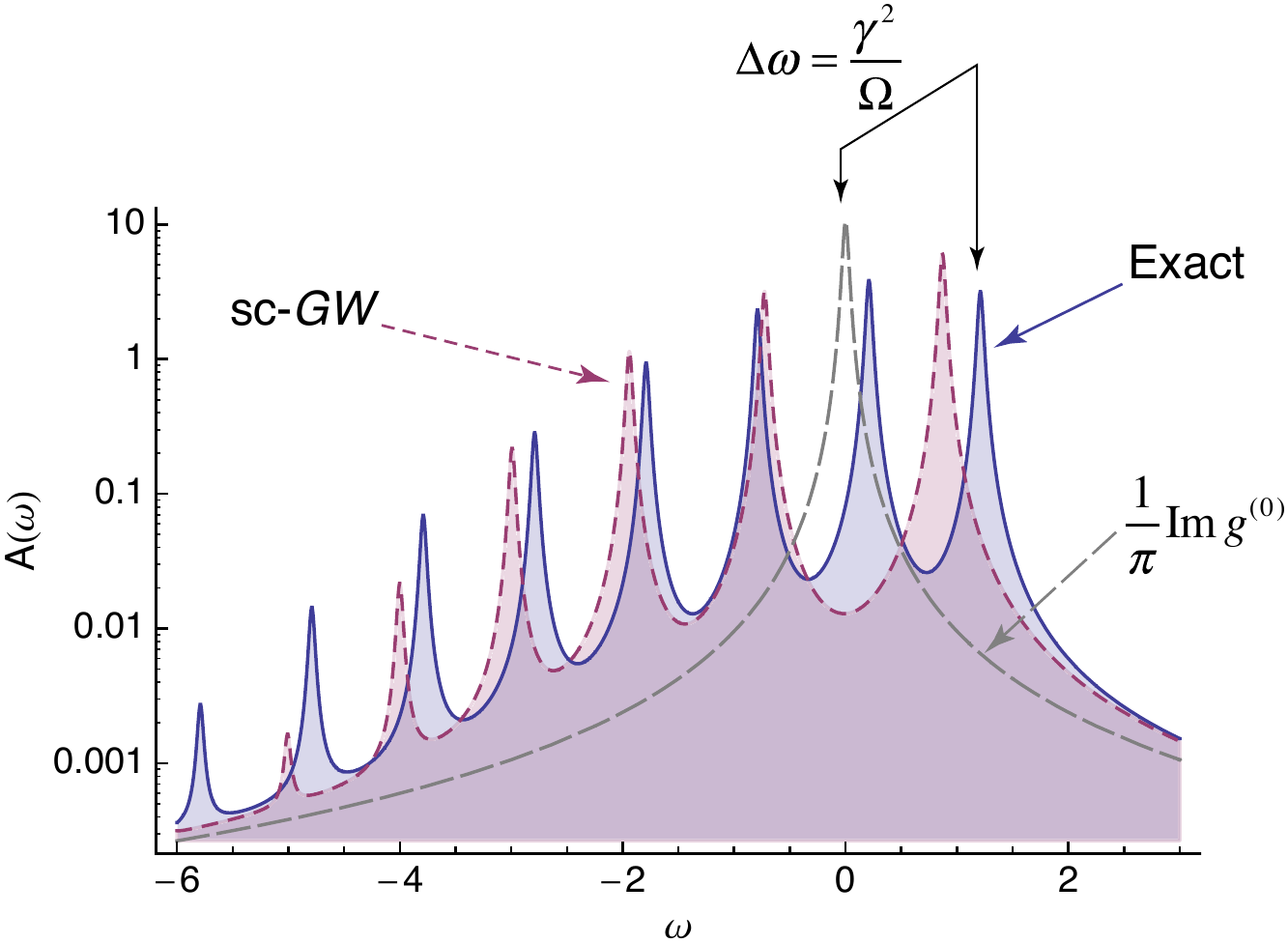}
\caption{(Color online) Spectral function at following values of parameters $\epsilon=0$,
  $\Omega=1$, $\gamma=1.1$, $\eta=0.03$ and different levels of theory: exact (full line),
  self-consistent first-order (short dashes), zeroth iteration (long dashes). Shaded areas
  are equal to unity.
\label{fig:A1}}
\end{figure}

Let us consider the following Green's function
\[
g(t-t')=-i\langle\psi|T [c(t)c^\dagger(t')]|\psi\rangle,
\]
where $|\psi\rangle$ is the exact ground state of the \emph{no-hole system}.  It can be
diagrammatically found by writing the cumulant expansion for the Green's function
$g(t)=g^{(0)}(t)e^{C(t)}$.  Observing that only a single diagram contributes to the
\emph{cumulant function} results in \be
C(t)=-\left(\frac{\gamma}{\Omega}\right)^2(1+i\Omega t-e^{i\Omega t}).
\label{eq:Ct}
\ee Corresponding exact spectral function is depicted at Fig.~\ref{fig:A1} together with
the zeroth order and spectral function from the self-consistent $GW$ calculation
(sc-$GW$).
The results are plotted for a strongly correlated regime ($\gamma>\Omega$) and can be
characterized as follows: (i) The spectral function consist of a main peak shifted by the
energy $\Delta \omega=\frac{\gamma^2}{\Omega}$ compared to the noninteracting case; (ii)
the quasiparticle peak is followed by the ladder of plasmonic satellites; (iii) the
self-consistent $GW$ method predicts the satellites. However, the position of even the
main peak is wrong. This inaccuracy is the main motivation for performing higher-order
diagrammatic calculations.

\begin{figure}[b!]
\includegraphics[width=\columnwidth]{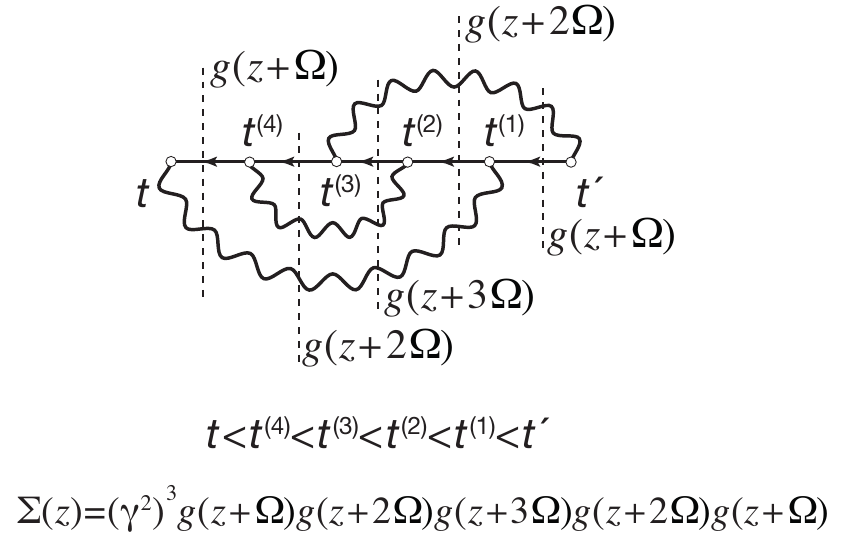}
\caption{Example of the self-energy in time domain. The system only contains
  holes. Therefore there is only one possible time-ordering as shown below the
  diagram. Bosonic propagators are denoted as wavy-lines. 
\label{fig:Diag_t}}
\end{figure}

\paragraph*{Diagrammatic properties\hspace{-1em}}---
Because the ground state is a no-hole state $c^\dagger|\psi\rangle$ vanishes and, hence,
the non-interacting time-ordered Green's function only consists of the hole propagator:
$g^{(0)}(t-t')=i\theta(t'-t)e^{-i\epsilon(t-t')}$.  This fact simplifies the diagrams
considerably: (i) in the expansion for the Green's function ($g$) and the self-energy
($\Sigma$) all intermediate points are time-ordered (Fig.~\ref{fig:Diag_t}); (ii) diagrams
containing loops necessarily yield a zero contribution. These properties allow to write
the self-energy for this model explicitly. Because there is no spatial degrees of freedom
the problem is similar to that of the Feynman diagrams enumeration which can be solved by
collapsing the space-time variables to one point (the zero-dimensional
model~\cite{molinari_hedins_2005,pavlyukh_analytic_2007}). Already such simplified model
has interesting applications for correlated electronic calculations in realistic
systems~\cite{lani_approximations_2012,berger_solution_2014,stan_multiple_2015}.  Here, we
present an analytic solution of a more complicated one-dimensional case.

Let $\Sigma^{(n,\alpha)}(\omega)$ be an $n$th-order self-energy term corresponding to a
particular diagram which will be denoted as $\alpha$. We will prove below that the
corresponding expression in the frequency representation is given by the product:
\be
\Sigma^{(n,\alpha)}[g;\omega]=(\gamma^2)^n\prod_{i=1}^{2n-1}g(\omega+k_i^{(n,\alpha)}\Omega),
\label{eq:Sigma_gen}
\ee 
where the integer number of absorbed plasmons in each fermionic line ($k_i^{(n,\alpha)}$)
is computed as a number of bosonic lines crossing each vertical line
(Fig.~\ref{fig:Diag_t}). $2n-1$ vertical lines are positioned such that they cut each
fermionic line. This equation can be derived by using the nonequilibrium Green's function
formalism. Let a vertical line separate times lying on the forward and backward branches
of the Keldysh contour in an expression for the \emph{lesser} self-energy
($\Sigma^<$). Consider, for instance, a third vertical line at Fig.~\ref{fig:Diag_t}. It
contributes $g^<(z-y_1-y_2-y_3)W^<(y_1)W^<(y_2)W^<(y_3)$ to $\Sigma^<$. Here,
$W^<(y)=\gamma^2\delta(y+\Omega)$ is the lesser bosonic propagator. Performing three
frequency integrals (over $y_1$, $y_2$, $y_3$) a contribution proportional to
$g^<(z+3\Omega)$ is obtained. Similar considerations can be repeated for each vertical
line and fermionic propagator yielding in total $2n-1$ terms for each $n$th-order
self-energy diagram $\Sigma^<(z)=\sum_{i=1}^{2n-1} f_i(z) g^<(z+k_i\Omega)$. Now, since
$f_i(z)$ are non-singular the generic expression for the time-ordered
self-energy~(\ref{eq:Sigma_gen}) is obtained.

Expansion (\ref{eq:Sigma_gen}) is a new exact result for the $S$-model which also permits
generalizations for more general scenarios. Electronic spectra of numerous realistic
materials have been rationalized in terms of the
time-ordered~\cite{aryasetiawan_multiple_1996,holm_self-consistent_1997,guzzo_valence_2011,pavlyukh_time_2013}
or retarded~\cite{kas_cumulant_2014} cumulant expansions, which as we have seen above, are
exact for the considered model. The presence of multiple plasmonic satellites is a marked
feature of these
materials~\cite{lischner_physical_2013,guzzo_multiple_2014,lischner_satellite_2014,kas_cumulant_2014,kas_real-time_2015}. The
plasmon dispersion is the only modification needed for generalization to this case. It
amounts to introducing additional sums over the plasmonic momentum at each vertex, but
does not change the diagrammatic structure. A viable route to use present results for the
momentum-resolved calculations is via the GF momentum average
approximation~\cite{berciu_systematic_2007}. It was demostrated to yield an accurate
description of dressed particles in the Holstein polaron model~\cite{berciu_greens_2006}.

\begin{figure}[t!]
\includegraphics[width=\columnwidth]{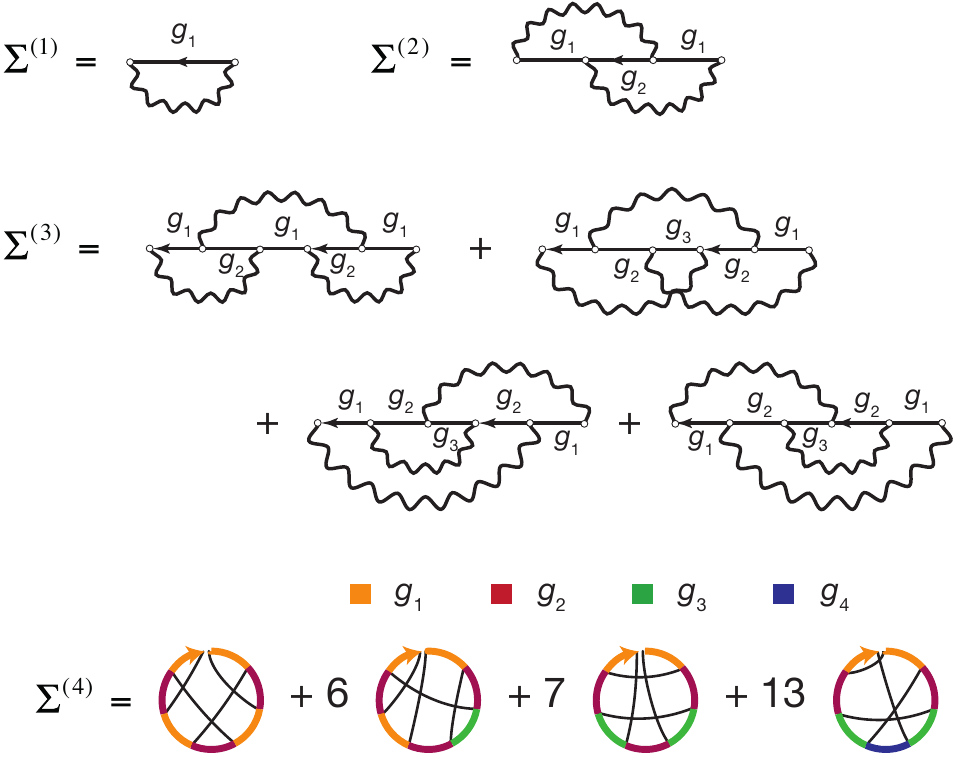}
\caption{Four lowest orders of the diagrammatic expansion of the self-energy for the
  $S$-model in frequency space. Notice that two diagrams of the third-order containing
  loops are not shown because they are equal to zero. The fourth-order self-energy is given
  in terms of chord diagrams with color-coding. Only one representative for each class is
  shown. Due to the absence of loops an isomorphism between the Feynman diagrams and the
  chord diagrams can be established.
\label{fig:Diag_2_3}}
\end{figure}

Eq.~(\ref{eq:Sigma_gen}) serves as the starting point for numerics; complexity goes into
the generation of Feynman diagrams and determination of the coefficients
$k_i^{(n,\alpha)}$. This is the second important ingredient of our approach. The
coefficients are computed purely algebraically by computing variational derivatives of
fermionic and bosonic propagators~\cite{strinati_application_1988} with respect to the
external position and time-dependent potential $\phi(1)$ (for simplicity time variables
are denoted as $t_i\equiv i$). $\Sigma[G,W]$ can be obtained by iterating a set of the
Hedin's equations~\cite{hedin_new_1965}. As was shown above the bosonic propagator in the
present model does not renormalize (loops give zero contribution), i.\,e. $\frac{\delta
  W(12)}{\delta \phi(3)}=0$, leading to a simpler set of equations:
\begin{subequations}
\label{eq:var_hedin}
\begin{align}
\Gamma(12,3)&=\delta(12)\delta(13)+\frac{\delta \Sigma(12)}{\delta V(3)},\\
\Sigma(12)&=i\int W(13)G(14)\Gamma(42,3)d(34),\\
\frac{\delta G(12)}{\delta V(3)}&=\int G(14)G(52)\Gamma(45,3)d(45),\label{eq:dgdv}
\end{align}
\end{subequations}
where $\Gamma(12,3)$ is the vertex function, $\Sigma(12)$ is the electron self-energy, and
$V(3)$ is the external [$\phi(3)$] plus the induced field in the system.  All these
quantities are functionally dependent on the external field $\phi(3)$ and on the full
electron propagator $G(12)$. The set of equations~(\ref{eq:var_hedin}) can now be iterated
starting from $\Gamma^{0}(12,3)=\delta(12)\delta(13)$ leading to the diagrams shown at
Fig.~\ref{fig:Diag_2_3}.

The chord diagram~\cite{pilaud_analytic_2014,hinich_cyclic_2002} representation is natural
in this case because according to the analysis above the fermionic loops yield zero
contribution. In order to further facilitate the interpretation of the graphs in frequency
space we use color coding for the coefficients $k_i^{(n,\alpha)}$ entering the GF
arguments. The graphs were generated by our symbolic algorithm in {\sc mathematica}
computer algebra system. Conversion from the time to frequency domains is likewise
performed using a symbolic algorithm. The self-energy accurate to the sixth order
comprises 1, 1, 4, 27, 248, and 2830 diagrams of the first to sixth orders, respectively,
and has the following algebraic representation:
\begin{align}
\Sigma&=\alpha  g_1+\alpha ^2 g_2 g_1^2+\alpha ^3 \left(g_2^2 g_1^3+3 g_2^2 g_3 g_1^2\right)\nn\\
&+\alpha ^4 \left(g_2^3 g_1^4+6 g_2^3 g_3 g_1^3+7 g_2^3 g_3^2 g_1^2+13 g_2^2 g_3^2 g_4 g_1^2\right)\nn\\
&+\alpha ^5 \big(g_2^4 g_1^5+9 g_2^4 g_3 g_1^4+23 g_2^4 g_3^2 g_1^3+26 g_2^3 g_3^2 g_4 g_1^3\nn\\
&+15 g_2^4 g_3^3 g_1^2+58 g_2^3 g_3^3 g_4 g_1^2+45 g_2^2 g_3^3 g_4^2 g_1^2+71 g_2^2 g_3^2 g_4^2 g_5 g_1^2\big)\nn\\
&+\alpha^6 \big(g_1^6 g_2^5+12 g_1^5 g_2^5 g_3+48 g_1^4 g_2^5 g_3^2+72 g_1^3 g_2^5 g_3^3\nn\\
&+31 g_1^2 g_2^5 g_3^4+39 g_1^4 g_2^4 g_3^2 g_4+194 g_1^3 g_2^4 g_3^3 g_4+183 g_1^2 g_2^4 g_3^4 g_4\nn\\
&+90 g_1^3 g_2^3 g_3^3 g_4^2+313 g_1^2 g_2^3 g_3^4 g_4^2+145 g_1^2 g_2^2 g_3^4 g_4^3\nn\\
&+142 g_1^3 g_2^3 g_3^2 g_4^2 g_5+310 g_1^2 g_2^3 g_3^3 g_4^2 g_5+470 g_1^2 g_2^2 g_3^3 g_4^3 g_5\nn\\
&+319 g_1^2 g_2^2 g_3^2 g_4^3 g_5^2+461 g_1^2 g_2^2 g_3^2 g_4^2 g_5^2 g_6\big)+\mathcal{O}(\alpha^7),
\label{S:6}
\end{align}
where $g_k\equiv g(\omega+k\Omega)$ and $\alpha\equiv\gamma^2$. Setting all $g_k\equiv1$ a
generating function for the enumeration of all chord diagrams is obtained:
\[
y(\alpha)=\alpha+\alpha ^2+4 \alpha ^3+27 \alpha ^4+248 \alpha ^5+2830 \alpha ^6+\mathcal{O}(\alpha^7),
\]
which also fulfills the following ordinary differential equation $2\alpha^2y\frac{d
  y}{d\alpha}+\alpha y^2-y+1=0$, resulting from (\ref{eq:var_hedin}) by collapsing all
time variables to one point~\cite{pavlyukh_analytic_2007}.

Our explicit form for the self-energy dictates that the singularities of $\Sigma$ should
be located exactly at the Green's function poles. Physically it is wrong as it is well
known that the self-energy poles lie between the poles of the corresponding exact Green's
function~\cite{winter_study_1972}. These two facts can be reconciled noticing that already
starting with the second order
\[
\Sigma^{(2)}(\omega)=(\gamma^2)^2g(\omega+\Omega)g(\omega+2\Omega)g(\omega+\Omega)
\]
the self-energy contains \emph{higher-order} poles in the frequency domain.  As in our toy
model (Fig.~\ref{fig:Pade}) they are responsible for the energy shift.

\paragraph*{Application to the $S$-model\hspace{-1em}}---Assume
\begin{figure}[t!]
\includegraphics[width=\columnwidth]{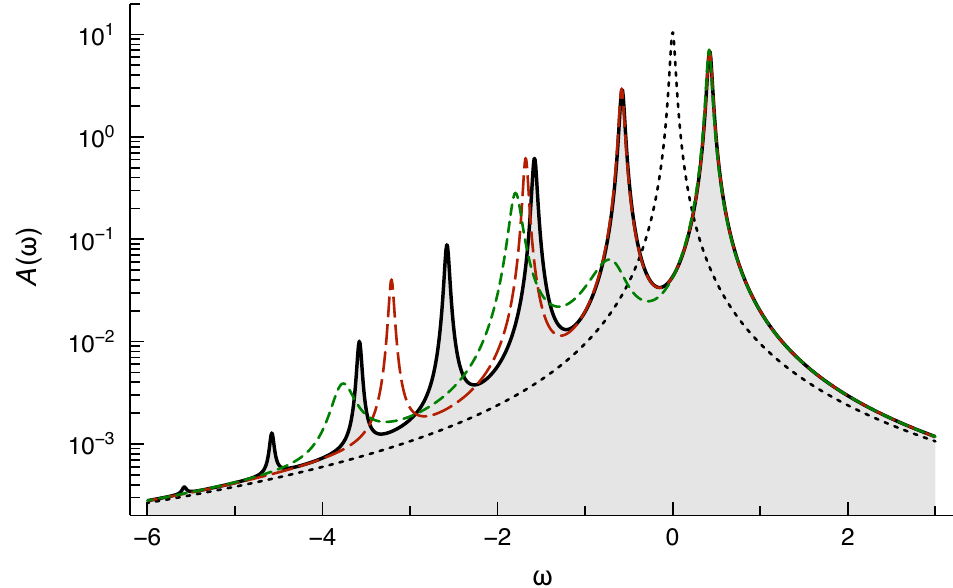}
\caption{Spectral function of the $S$-model at different levels of theory: exact (full
  line), self-consistent third-order (short dashed), sixth-order (long dashed), zeroth
  iteration (dotted) for the following values of parameters: $\epsilon=0$, $\Omega=1$,
  $\gamma=0.65$, $\eta=0.03$.
\label{fig:A6}}
\end{figure}
 that in the course of a self-consistent calculation an approximation for the Green's
 function has been obtained ($g^{(i)}(\omega)$). Using the diagrammatic expansion
 viz. Eq.~(\ref{S:6}) we compute an approximation to the self-energy
 $\Sigma[g^{(i)}](\omega^*)$ at a chosen frequency point. The point $\omega^*$ should
 belong to the domain where the perturbative expansion converges. In order to obtain the
 self-energy in the vicinity of Green's function poles where the series diverge we perform
 the Pad{\'e} approximation
 $\Sigma[g^{(i)}](\omega^*)\rightarrow\tilde{\Sigma}^{(i)}(\omega)$ and use the new
 self-energy in order to update the Green's function according to the Dyson equation
 $g^{(i+1)}(\omega)=[\omega-\epsilon-\tilde{\Sigma}^{(i)}(\omega)]^{-1}. $ Iterations are
 started from the noninteracting GF $g^{(0)}(\omega)=(\omega-\epsilon-i\eta)^{-1}$ and
 typically converge within some tens of cycles. The quality of the resulting spectral
 function strongly depends on the order of perturbative expansions and on the strength of
 the electron-plasmon scattering $\gamma$. For the weakly correlated regime $\gamma\simeq
 0.1\Omega$ already the $GW$ approximation faithfully reproduces the exact spectral
 function. This approximation ceases to be valid in the \emph{correlated regime} as
 Fig.~\ref{fig:A1} demonstrates. The energy of the main quasiparticle (QP) peak is the
 major discrepancy.  However, for $\gamma=0.65$ already third-order treatment yields very
 good results for QP energy and strength (Fig.~\ref{fig:A6}, short dashed line). The first
 satellite, which has a rather large contribution to the density of states at this value
 of $\gamma$ (notice logarithmic scale), represents a substantially more complicated
 feature. It can only be captured with a self-energy that is accurate to the 6th order
 (long dashed line). However, even 3111 diagrams are not sufficient to reproduce the
 second-order satellite! For realistic systems methods based on the notion of four-point
 vertex $\Gamma$~\cite{blaizot_quantum_1986,van_leeuwen_total_2006} might be a viable
 alternative. In full generality $\Gamma(12,34)$ can be obtained by solving a set of
 coupled Bethe-Salpeter equations in particle-particle ([12]) and particle-hole ([14] and
 [13]) channels known as parquet
 equations~\cite{roulet_singularities_1969,gogolin_bosonization_1998,rohringer_local_2012}. If
 only one such channel is considered one arrives at the so-called $T$-matrix
 approximations (TMA)~\cite{romaniello_beyond_2012,peng_equivalence_2013} known to
 complement the
 $GW$-approximation~\cite{nagano_correlations_1984,verdozzi_evaluation_1995,nechaev_variational_2005,qian_-top_2006,
   von_friesen_successes_2009,yang_benchmark_2013}.  We have verified that omitting 22 and
 714 diagrams of the 5th and 6th orders from Eq.~(\ref{S:6}) according to the parquet
 procedure with only one simple four-vertex does not detiorate the quality of the
 results. Even better description of higher order satellites can be expected if the
 parquet procedure is iterated further.

\paragraph*{Conclusions\hspace{-1em}}--- 
It is more than computational complexity that prevents applications of many-body
perturbation theory beyond the leading order. Resulting asymptotic series lead to Green's
functions with incorrect physical properties: non-positive densities, higher-order poles
already for the second
order~\cite{stefanucci_diagrammatic_2014,uimonen_diagrammatic_2015}. For various
statistical models the Pad{\'e} approximation has been used to extend perturbative
expansions beyond their domain of convergence~\cite{kleinert_critical_2001}. The same
mathematical approach is used here in a different context, to regularize the electron
self-energy.  With the help of nonequilibrium Green's function theory we have derived the
self-energy of the $S$-model explicitly and demonstrated a connection of its diagrammatic
expansion to a certain class of chord diagrams. With the help of the developed symbolic
algorithm analytical expressions up to the sixth order in the electron-plasmon interaction
are generated. For $\omega^*\gg\frac{\gamma^2}{\Omega}$ the series converge rapidly,
however, there are no interesting spectral features in this domain. Therefore, to
recursively update the Green's function in the whole spectral range the self-energy is
regularized before plugging it into the Dyson equation. In this way, even in the
correlated regime ($\gamma=0.65\Omega$) the present approach allows to accurately describe
the QP peak and the first-order satellite. Hence, the Pad{\'e} approximation makes
self-consistent calculations with higher order vertex function feasible.
\begin{acknowledgments}
YP and JB acknowledge financial support of the German Research Foundation (DFG) through
SFB 762 and PA 1698/1-1 projects.  AR acknowledges financial support from the European
Research Council Advanced Grant DYNamo (ERC-2010- AdG-267374), Spanish Grant
(FIS2013-46159-C3-1-P), Grupos Consolidados UPV/EHU del Gobierno Vasco (IT578-13) and
European Community FP7 project CRONOS (Grant number 280879-2) and COST Actions CM1204
(XLIC) and MP1306 (EUSpec).
\end{acknowledgments}
\end{document}